\documentclass[%
 reprint,
 amsmath,amssymb,
 aps,
prd,
]{revtex4-2}
\usepackage{xspace}
\usepackage{graphicx}
\usepackage{dcolumn}
\usepackage{natbib}
\usepackage{bm}
\usepackage{xcolor}
\usepackage{lineno}

\usepackage[colorlinks=true,linkcolor=blue,citecolor=blue,urlcolor=blue]{hyperref}

\begin{document}

\preprint{APS/123-QED}

\title{A White Dwarf Tidal Disruption by an Intermediate-Mass Black Hole as the Progenitor of Ultra-long GRB 250702B}

\author{Chengchao Yuan$^1$}\email[Contact author: ]{chengchao.yuan@ulb.be}
\author{Ning Jiang$^{2,3}$}%
\author{Zi-Gao Dai$^{2,3}$}

\affiliation{$^1$Université Libre de Bruxelles, CP225 Boulevard du Triomphe, 1050 Brussels, Belgium}%
\affiliation{$^2$Department of Astronomy, University of Science and Technology of China, Hefei, 230026, China}
\affiliation{$^3$School of Astronomy and Space Sciences,
University of Science and Technology of China, Hefei, 230026, China}

\date{\today}

\begin{abstract}       
The recent detection of GRB 250702B, the longest gamma-ray burst observed to date with prompt emission lasting $\sim 2.5\times 10^4$ seconds, challenges the conventional collapsar model. Its remarkable features—including an extraordinary X-ray flare at $\sim 1.3$ days post-detection, a late-time transition from steep to shallow decay in the X-ray afterglow, and hard spectra extending from keV to MeV energies—point to a novel progenitor. Here we show that these multiwavelength signatures can be consistently explained by a relativistic jet powered by successive partial tidal disruptions of a white dwarf (WD) by an intermediate-mass black hole (IMBH). By modeling the time-dependent accretion rate from repeated partial disruptions and the resulting jet evolution, we show that the external forward and reverse shocks account for the long-term X-ray, near-infrared, and radio afterglow, whereas the luminous X-ray flare originates from internal energy dissipation caused by collisions between fast and slow relativistic ejecta associated with the final complete disruption. Our findings establish IMBH–WD tidal disruption events as a viable engine for ultra-long GRBs.


\end{abstract}

\maketitle

\section{Introduction}\label{sec:intro}
Gamma-ray bursts (GRBs) are among the most luminous transients in the Universe and are widely understood as radiation from collimated, relativistic jets launched by compact central engines (e.g., Refs. \cite{Sari:2000zp,Piran:2004ba,Meszaros:2006rc}). Observationally, GRBs are commonly classified into short and long categories according to their prompt duration (e.g., Ref. \cite{Kouveliotou:1993yx}), with short bursts primarily associated with compact object mergers and long bursts linked to the collapse of massive stars (e.g., Refs. \cite{Woosley:2006fn,Nakar:2007yr,Berger:2013jza}). In both cases, dissipation within the relativistic outflow produces the prompt gamma-ray emission, while the subsequent interaction of the ejecta with the ambient medium generates broadband afterglow emission through forward and reverse shocks (e.g., Refs. \cite{Kumar:2014upa,2018pgrb.book.....Z}). Because the prompt duration broadly reflects the activity timescale of the central engine, it is typically limited to $\lesssim10^3$ s in collapsar models by the accretion time of the stellar envelope.

A rare subclass, known as ultra-long GRBs \cite{Levan:2013gcz}, exhibits prompt emission lasting $\gtrsim 10^3$ s and in some cases extending to several hours, challenging the standard collapsar paradigm. Notably, the recently discovered GRB 250702B (redshift $z \simeq 1.036$ \cite{Gompertz:2025wcq}) displayed gamma-ray activity persisting for $\sim 2.5 \times 10^4$ s, making it the longest GRB detected to date \cite{Neights:2025keq,Li:2025mae}, while retaining key signatures of relativistic jets, such as hard spectra, rapid variability, and large energetics, thereby motivating alternative interpretations involving extended accretion or sustained mass supply to the central engine.

Multiwavelength observations have provided important insight into the nature of this event. The transient was first discovered in X-rays by the \emph{Einstein Probe} (EP) on 2 July 2025 \cite{Yuan:2022fpj,2025GCN.40906....1C} and subsequently by the \emph{Fermi} Gamma-ray Burst Monitor (GBM) \cite{2025GCN.40891....1N,2025GCN.40931....1N}, Monitor of All-sky X-ray
Image (MAXI) \cite{2025GCN.40910....1K}, Konus-Wind \cite{2025GCN.40914....1F}, and the Space Variable Objects Monitor (SVOM) \cite{2025GCN.40923....1S}, which jointly revealed bright, highly variable X-ray emission preceding and accompanying the gamma-ray activity such as recurring flares \cite{Oganesyan:2025esg,Neights:2025keq,Li:2025mae,Zhang:2025xds}, while continued monitoring showed long-lived X-ray evolution over days to weeks. The long-lasting emission episode, together with the rapid temporal evolution observed in X-rays and gamma-rays, as well as follow-up detections in the radio, near-infrared, optical and X-ray bands \cite{Li:2025mae,Levan:2025mrt,Carney:2025dpi,OConnor:2025ubn}, favor a tidal disruption origin in which a white dwarf (WD) is disrupted by an intermediate-mass black hole (IMBH) \cite{Levan:2025mrt,Beniamini:2025xsk,Eyles-Ferris:2025rhy,Li:2025mae,Granot:2025pmw,Sato:2026mem}. Such IMBH–WD disruptions naturally yield short mass fallback times and high peak accretion rates, enabling the launch of highly beamed relativistic jets under strongly super-Eddington conditions (e.g., Refs. \cite{Burrows:2011dn,Bloom:2011xk,Andreoni:2022afu,Yuan:2024daj,Yuan:2024sxk}), thereby providing a plausible framework for the temporal and spectral behavior observed in GRB 250702B. For instance, Ref.~\citep{Sato:2026mem} suggested that the recurring gamma-ray flares could originate from successive partial IMBH–WD disruptions occurring when the remnant of the WD passes through the orbital pericenter for multiple times (e.g., Refs. \cite{Zalamea:2010mv,MacLeod:2014mha,Chen:2022oin,Lau:2025cmv,Chen:2025zpv}).

\begin{figure*} [htp]
    \centering
    \includegraphics[width=0.8\linewidth]{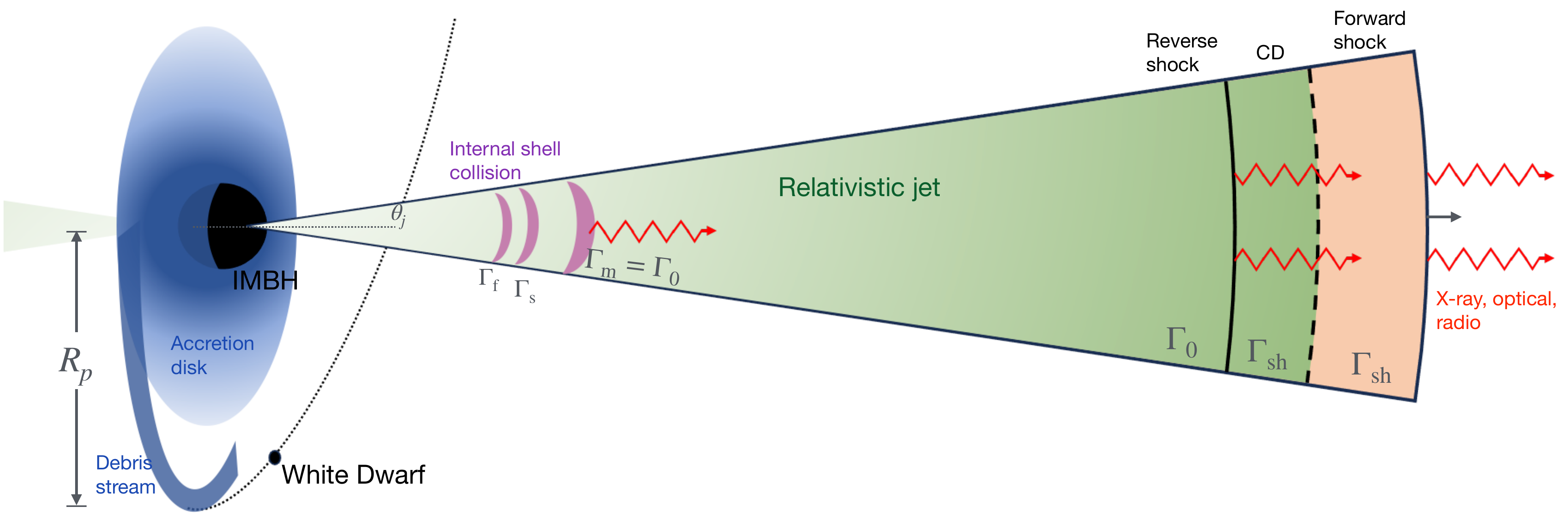}
    \caption{Schematic illustration of the partial disruption of a WD by an IMBH, along with the structure of a relativistic jet propagating through the external medium. The dotted curve represents the orbit of the WD, while the debris stream and the resulting accretion disk are also shown. The pericenter distance is denoted as $R_p$. Within the jet, the structures from right to left correspond to the forward shock, contact discontinuity, reverse shock, and internal shell collisions.}
    \label{fig:schematic}
\end{figure*}

In this work, we investigate the temporal and spectral signatures of a relativistic jet powered by successive partial IMBH–WD disruptions, incorporating persistent energy injection from accretion, the evolution of external forward and reverse shocks, and internal shocks produced by collisions between relativistic ejecta (referred to as `shells’ hereafter) with different Lorentz factors, which have been widely proposed as the origin of GRB prompt emission (e.g., Refs. \cite{Rees:1994nw,Kobayashi:1997jk,Daigne:1998xc,Rudolph:2022ppp}). The analysis focuses on these prominent observational features \cite{Li:2025mae}: (i) an extraordinary X-ray flare occurring $\sim 1.3$ d after the first observation by EP, with a flux $\sim 30–100$ times higher than the afterglow baseline and a duration of $\sim 0.3$ d in the observer's frame; (ii) the late-time transition of the X-ray light curve from steep to shallow decay; and (iii) the hard spectra extending from keV to MeV energies in both the X-ray flaring and afterglow phases.

Specifically, we derive the time evolution of the accretion rate onto the IMBH from repeated partial disruptions in \S\ref{sec:disruption}, and subsequently model the dynamics of the relativistic jet propagating through an external medium with continuous power injection (\S\ref{sec:model}), along with the properties of the external and internal shock regions. This framework enables time-dependent modeling of multiwavelength emission within a synchrotron self-Compton scenario, allowing us to reproduce the observed spectra and light curves. A summary and discussion are provided in \S\ref{sec:discussion}. Throughout, primed symbols denote quantities in the jet comoving frame, with $T$, $t$, and $t'$ representing the observer, IMBH-rest, and jet-comoving times, respectively.

\section{Accretion rate of IMBH-WD partial disruptions}\label{sec:disruption}
Considering a WD of mass $M$ passing by an IMBH of mass $M_{\rm BH}\sim10^{3}$–$10^5~M_\odot$ on an eccentric orbit, a small pericenter distance can lead to multiple partial tidal disruptions of the WD before a final catastrophic disruption occurs, as illustrated schematically in Figure~\ref{fig:schematic}. The fraction of WD mass that is tidally stripped during each pericenter passage can be written as \cite{Zalamea:2010mv}
\begin{equation}
\frac{\delta M}{M}=6.1\left(1-\frac{M}{ M_{\rm Ch}}\right)^{3\beta/2}\left(\frac{\Delta}{R}\right)^{5/2},
\label{eq:mass_loss}
\end{equation}
where $M_{\rm Ch}\simeq1.43M_\odot$ is the Chandrasekhar mass, and $\beta\simeq0.447$ provides a good description of the mass loss for $0.2M_\odot\lesssim M\lesssim1.4M_\odot$. In this expression, the ratio of the width of the stripped surface shell ($\Delta$) to the WD radius ($R$) can be explicitly written as
\begin{equation}
\frac{\Delta }{R}=1-\left(\frac{R_p}{2R_T}\right)\left(\frac{M}{M_0}\right)^{2/3}\left(\frac{M_{\rm Ch}-M_0}{M_{\rm Ch}-M}\right)^\beta,
\label{eq:shell_width}
\end{equation}
where $R_T=R(M_{\rm BH}/M)^{1/3}$ is the tidal radius, $R_p=a(1-e)$ is the pericenter distance in terms of the semimajor axis $a$ and eccentricity $e$, and $M_0$ is the initial WD mass before the first disruption. Eq.~\ref{eq:mass_loss} automatically accounts for the change in the WD radius as $M$ decreases, e.g.,
\begin{equation}
    R \approx 0.013R_\odot\left(\frac{M_{\rm Ch}}{M}\right)^{1/3}\left(1-\frac{M}{M_{\rm Ch}}\right)^{\beta}.
\end{equation}
We note that the changes in $a$ and $e$ due to gravitational radiation is insignificant, as the following conditions are satisfied:
\begin{equation}
\frac{P}{a}\left|\frac{da}{dt}\right|\ll1, \qquad \frac{P}{e}\left|\frac{de}{dt}\right|\ll1,
\end{equation}
where the expressions for $da/dt$ and $de/dt$ are adopted from Ref. \cite{Peters:1964zz}.

As $M$, $R_T$, and $R_p$ evolve over time, a final complete disruption occurs when either $R_p<R_T$ or $\delta M/M>1$ is satisfied. Denoting the total number of disruptions (including the complete disruption) as $N_{\rm dis}$, and assuming that each stripped mass element $\delta M$ independently falls back onto the IMBH with a rate $\propto t^{-5/3}$, we derive the accretion rate as a function of time by summing the contributions from the $i$-th passage,
\begin{equation}
\dot M_{\rm BH}(t)={\eta_{\rm acc}}\sum_{i=1}^{N_{\rm dis}}\frac{\delta M_{i}}{3P_{i}}\left(\frac{t}{P_{i}}\right)^{-5/3}\Theta\left(t-\tilde t_{i}\right),
\label{eq:acc_rate}
\end{equation}
where $\eta_{\rm acc}$ represents the fraction of the stripped mass that is ultimately accreted. In this formulation, the starting time $t=0$ is defined as the time of the first disruption, $P=2\pi\sqrt{a^3/GM_{\rm BH}}$ is the orbital period of both the debris and the WD remnant, and $\tilde t_{i}=\sum_{j=1}^{i-1}P_{j}$ with $\tilde t_{1}=0$ denotes the time of the $i$-th disruption. The function $\Theta(x)$ is defined such that $\Theta(x)=1$ for $x>0$ and $\Theta(x)=0$ otherwise.

\begin{figure}
    \centering
    \includegraphics[width=1.0\linewidth]{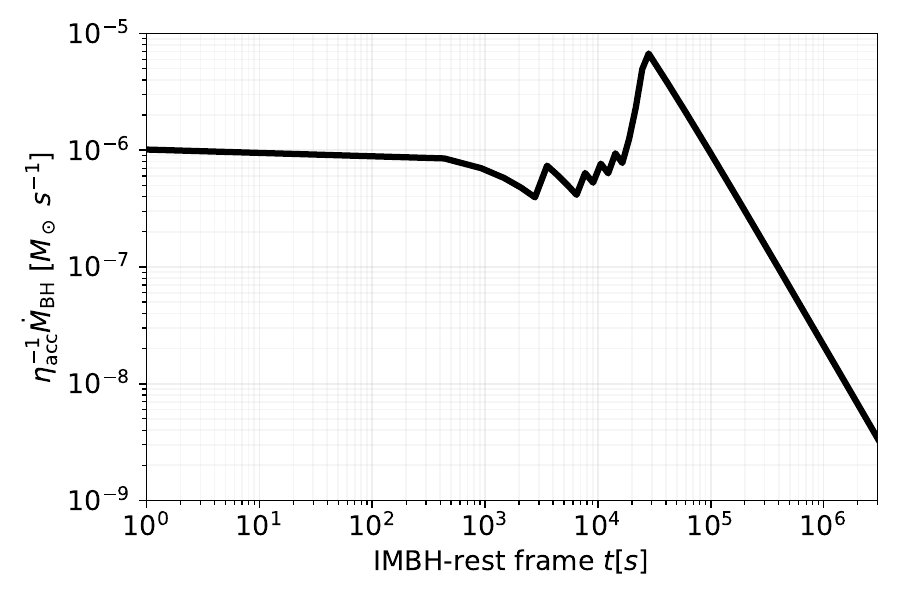}
    \caption{Accretion rate (defined by Eq. \ref{eq:acc_rate}) multiplied by $\eta_{\rm acc}^{-1}$ as a function of time during the successive partial disruptions of the WD and following the final complete disruption (e.g., $\dot M_{\rm BH}\propto t^{-5/3}$ for $t\gtrsim3\times10^{4}~\rm s$). The reference time is defined as the time of the first disruption.}
    \label{fig:acc_rate}
\end{figure}

We fix $M_{\rm BH}=10^4M_\odot$, and adopt $M_0 = 1.0M_\odot$, $R_p=1.8R_T$ and $e=0.97$ as the initial parameters, such that the time of final disruption is comparable to the X-ray flare occurring $\sim 1$ day after the first observation by EP. Figure~\ref{fig:acc_rate} depicts the accretion rate (multiplied by $\eta_{\rm acc}^{-1}$) as a function of time measured in the IMBH rest frame. In this configuration, the WD undergoes a total of $N_{\rm dis}=9$ partial disruptions with an orbital period $P\sim{\rm a~few}\times10^{3}~\mathrm{s}$. The peak accretion rate at $t\sim3\times10^{4}~\mathrm{s}$ is produced jointly by the pile-up effect of the final multiple disruptions and by the last, complete disruption. 

\begin{figure*}
    \centering
    \includegraphics[width=0.49\textwidth]{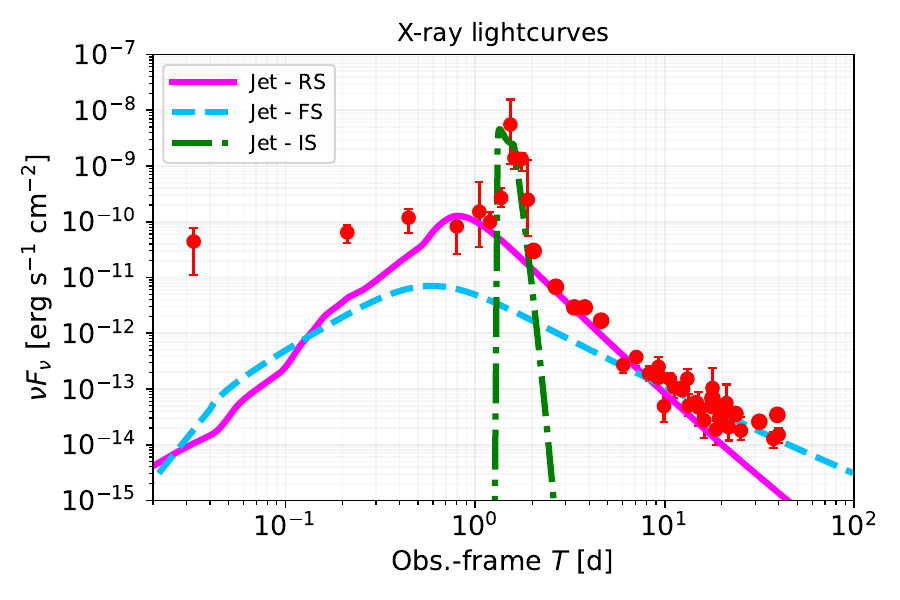}
    \includegraphics[width=0.49\textwidth]{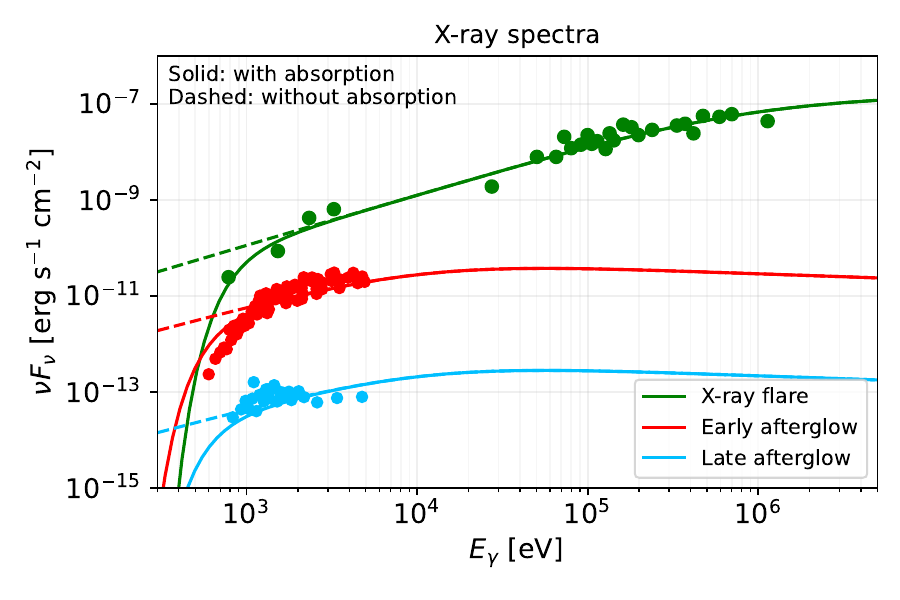}

    \caption{X-ray light curves ($0.3-10$ keV band, left panel) and spectra (right panel) of GRB 250702B. The data points show observations by EP \citep{Li:2025mae}. In the left panel, the magenta solid, blue dashed, and green dash-dotted curves depict the contributions from the jet reverse shock, forward shock, and internal energy dissipation regions, respectively. In the right panel, the green, red, and blue points illustrate the X-ray spectra during the X-ray flare stage, early afterglow (EP-FXT follow-up observation on 3 July 2025, $\sim2$ days after the trigger), and late-time stacked spectrum (data taken from 12 July to 15 July), respectively. The dashed and solid curves show the corresponding spectra predicted by the TDE jet model before and after accounting for absorption.}
    \label{fig:lc_spec}
\end{figure*}

\section{Jet signatures}\label{sec:model}
\subsection{Jet structure}

Similar to jetted tidal disruption events (TDEs), such as AT~2022cmc \cite{Andreoni:2022afu,Pasham:2022oee}, Swift~J1644 \cite{Bloom:2011xk,Burrows:2011dn,Levan:2011yr,Zauderer:2011mf}, Swift~J2058 \cite{Cenko:2011ys,Pasham:2015bua}, and Swift~J1112 \cite{Brown:2015amy,Brown:2017iuv}, where relativistic jets are invoked to interpret the afterglow emission in the X-ray and radio bands \citep{Zauderer:2011mf,2011MNRAS.416.2102G,2012MNRAS.420.3528M,2012ApJ...748...36B,2013ApJ...767..152Z,2018ApJ...854...86E,2021ApJ...908..125C,2023MNRAS.522.4028M, 2024ApJ...963...66Z,Sato:2024zpq, Yuan:2024daj,Yuan:2024sxk}, we systematically investigate the emission from the jet launched after the initial IMBH–WD partial disruption. The physical picture is that, when the jet with an initial Lorentz factor $\Gamma_{0}$ propagates into the ambient gaseous medium, it sweeps up material, leading to the formation of a forward shock (FS) that accelerates the external medium (upstream) to a Lorentz factor $\Gamma_{\rm sh}\lesssim\Gamma_{0}$, and a reverse shock (RS) that decelerates the unshocked ejecta from $\Gamma_{0}$ to $\Gamma_{\rm sh}$. Figure~\ref{fig:schematic} schematically illustrates the geometry of the FS, RS, and the contact discontinuity (CD) separating their downstream regions.

We parameterize the isotropic-equivalent continuous power and mass injection into the jet using the time-dependent accretion rate (e.g., Eq.~\ref{eq:acc_rate}), respectively as $L_{j,\rm iso}=\eta_j\dot M_{\rm BH}c^2/(\theta_j^2/2)$ and $dM_{j,\rm iso}/dt=L_{j,\rm iso}/(\Gamma_0c^2)$, where $\eta_j$ represents the fraction of accreted power reprocessed into the jet and $\theta_j$ is the jet opening angle. Following Ref.~\cite{Yuan:2024daj}, we solve the coupled differential equations governing the jet evolution, incorporating the mass and energy transfer between the FS and RS, the continuous injection, as well as a generic external density profile that connects the distance-dependent circumnuclear medium (CNM) of radius $R_{\rm cnm}$ to the interstellar medium (ISM),
\begin{linenomath*}
\begin{equation}
n_{\rm ext}(R_j)=
\begin{cases}
n_{\rm ISM}\left(\frac{R_j}{R_{{\rm cnm}}}\right)^{-k}, & R_j<R_{{\rm cnm}}\\
n_{\rm ISM}, & R_j>R_{{\rm cnm}},
\end{cases}
\label{eq:n_ext}
\end{equation}
\end{linenomath*}
where $R_{j}$ is the jet radius, $R_{\rm cnm}\sim10^{18}\mathrm{cm}$ is motivated by the characteristic radii of winds emanating from accretion disks before merging into the ISM, $n_{\rm ISM}$ is the number density of the ISM, and radio modeling of jetted TDEs typically yields $k\sim1.5$–$2.0$ (e.g., Refs. \cite{Metzger:2011qq,Berger:2011aa}). The outputs of such jet dynamical modeling, including the time dependence of $\Gamma_{\rm sh}$, the jet radius $R_j$ and the downstream densities of the FS and RS, enable us to compute the time-dependent multiwavelength emission from these regions, which is used to explain the decaying X-ray light curves (e.g., the red points in the left panels of Figure~\ref{fig:lc_spec}). A top-hat jet structure and an on-axis viewing assumption are adopted.

However, the external shock model presented above fails to reproduce the X-ray flare observed at $T\sim1$ d (see the left panel of Figure~\ref{fig:lc_spec}), during which the X-ray flux rises rapidly by a factor of $\sim30$–$100$ and subsequently declines to the afterglow baseline level on a timescale of $\Delta T_{\rm X,flare}\sim0.3$ d. This feature might be the reminiscent of prompt emission, implying possible internal energy dissipation temporally associated with the peak of the accretion rate shown in Figure~\ref{fig:acc_rate}. We therefore propose a phenomenological internal shock model and consider the flash emission from collisions of relativistic shells at the jet base (see Figure~\ref{fig:schematic}). 

\subsection{External shock radiation}\label{sec:external}
As Refs.~\cite{Yuan:2024daj,Yuan:2024sxk} have shown that the fast-decaying X-ray afterglow of jetted TDEs can be attributed to emission from the RS region, whereas the FS typically produces a more shallowly decaying afterglow, we employ the RS and FS components to fit the X-ray light curve in the time intervals $0.6~{\rm d}\lesssim T\lesssim10~{\rm d}$ and $T\gtrsim10~{\rm d}$, respectively, excluding the flare. 

For the forward shock region, the magnetic field strength can be parameterized as $B_{\rm fs}'=\left[32\pi\epsilon_B^{\rm fs}\Gamma_{\rm sh}(\Gamma_{\rm sh}-1)n_{\rm ext}m_pc^2\right]^{1/2}$, where $\epsilon_{B}^{\rm fs}$ denotes the fraction of internal energy density converted into magnetic fields. We adopt a power-law distribution for the injection of shock-accelerated non-thermal electrons, $\dot Q_{\rm fs}'\propto\gamma_e'^{-s}$, where $\gamma_e'$ is the electron Lorentz factor and $s$ is the spectral index. The normalization is obtained from $(4\pi R_j^2 ct_{\rm dyn}')\int Q_{\rm fs}'d\gamma_e'=(4\pi f_e^{\rm fs} R_j^3n_{\rm ext})/(3t_{\rm dyn}')$, with the comoving dynamical time $t_{\rm dyn}'=R_j/(\Gamma_{\rm sh}c)$ and the accelerated-electron fraction $f_e^{\rm fs}$. The minimum Lorentz factor is given by $\gamma_{\rm min,fs}'=(\Gamma_{\rm sh}-1)g(s)(\epsilon_e^{\rm fs}/f_{e}^{\rm fs})(m_p/m_e)$, where $g(s)=(s-2)/(s-1)$, and $\epsilon_e^{\rm fs}$ represents the fraction of internal energy density transferred to non-thermal electrons. We then compute the time-dependent synchrotron and synchrotron self-Compton (SSC) emission by numerically solving the electron transport equation using the AM$^3$ software \cite{Klinger:2023zzv}, taking into account adiabatic cooling. The jet-break correction factor $f_{\rm br}=1/[1+(\Gamma_{\rm sh}\theta_j)^{-2}]$ is applied, and the observed flux is obtained by integrating over the equal-arrival-time surface. 

Similarly, the reverse shock emission can be calculated using the corresponding microphysics parameters, $f_e^{\rm rs}$, $\epsilon_{e}^{\rm rs}$, and $\epsilon_B^{\rm rs}$, except that $\Gamma_{\rm sh}$ is replaced by the relative Lorentz factor between the upstream and downstream, $\Gamma_{\rm rs,rel}\approx(\Gamma_{\rm sh}/\Gamma_0+\Gamma_0/\Gamma_{\rm sh})/2$, when calculating the magnetic field strength and the minimum electron Lorentz factor. The normalization of the RS electron injection rate is given by $(4\pi R_j^2t_{\rm dyn}'c)\int \dot Q_{\rm rs}'d\gamma_{e}'=f_e^{\rm rs}L_{j,\rm iso}/(\Gamma_0m_pc^2)$. In both cases, the same spectral index $s$ is adopted, and the maximum electron Lorentz factors are determined by balancing the acceleration rate, e.g., $eB'/(\gamma_e' m_e c)$, with the cooling rates due to synchrotron and SSC processes.

The solid magenta and dashed blue curves in the left panel of Figure~\ref{fig:lc_spec} correspond to the X-ray light curves from the RS and FS regions, obtained with parameters (as summarized in Table~\ref{tab:params}) comparable to those used in modeling the X-ray afterglows of jetted TDEs \cite{Yuan:2024sxk}. The spectral index is fixed at $s=2.2$, and $k=2.0$ is adopted for the external density profile (Eq.~\ref{eq:n_ext}). The numerical results are consistent with analytical scalings in the post–jet-break regime \cite{Yuan:2024daj}, e.g., $\nu F_\nu\propto \Gamma_{\rm sh}^2T^{-(s+8)/6}$ and $\nu F_\nu\propto \Gamma_{\rm sh}^{2}T^{-1}$ for the RS and FS, respectively, with $\Gamma_{\rm sh}\propto T^{-3/8}$ for $T>(1+z)\tilde t_{N_{\rm dis}-1}$. Hence, the transition of the X-ray afterglow from a steep to a shallow decay at $T\sim10$ d is naturally interpreted as a shift from RS to FS dominance. In addition, the red and blue curves in the right panel of Figure~\ref{fig:lc_spec} illustrate the X-ray spectral fits for the early observations by the EP Follow-up X-ray Telescope (FXT) and the late-time stacked spectrum \cite{Li:2025mae}, respectively. After correcting the Galactic absorption with a hydrogen column density of $N_H\sim3.3\times10^{21}~\rm cm^{-2}$ and the intrinsic absorption with redshift $z=1.036$ using the Xspec package \cite{1996ASPC..101...17A}, the observed hard X-ray spectra can be well described by synchrotron radiation (solid curves), whereas the dashed curves represent the unabsorbed fluxes.

\begin{table}
\centering
\caption{Model parameters for the ambient gas medium, jet structure, and radiation regions ($k$ and $s$ are fixed to be $2.0$ and $2.2$, respectively). \label{tab:params}}
\begin{tabular}{l l}
\hline\hline
\textbf{Parameter} & \textbf{Value} \\
\hline
$n_{\rm ISM}$ & 2.0 cm$^{-3}$ \\
$\eta_{\rm j}\eta_{\rm acc}$ & 0.15 \\
$\theta_{\rm j}$ & 0.05 (2.9$^\circ$) \\
$\Gamma_{0}$ & 30 \\
\hline
$\epsilon_{e}^{\rm fs},~\epsilon_{e}^{\rm rs}$ & $0.02$, 0.03 \\
$\epsilon_{B}^{\rm fs},~\epsilon_{B}^{\rm rs}$ & $0.007$, 0.02 \\
$f_{e}^{\rm fs},~f_{e}^{\rm rs}$ & 0.1, ~$1.5\times10^{-3}$ \\
\hline
$T_{\rm coll}$, $\Delta T_{\rm coll}~~~~$ & 1.3 d, 0.3 d \\
$M_{\rm shell}$  & $10^{-3}~M_\odot$ \\
$\Gamma_{\rm f}$, $\Gamma_{\rm m}$ & 60, 15 \\
$\epsilon_{e}^{\rm is}$, $\epsilon_{B}^{\rm is}$ & 0.10, $1.2\times10^{-3}$ \\
$f_{e}^{\rm is}$ & $2.0\times10^{-3}$ \\
\hline\hline
\end{tabular}
\end{table}



\subsection{Internal energy dissipation via shell collisions}\label{sec:internal}
As demonstrated before, the external shock scenarios cannot reproduce the luminous X-ray flare at $T_{\rm flare}\sim1.3~\rm d$. We instead propose an internal shock model involving collisions of relativistic shells near the jet base, where a fast shell with mass $M_{\rm f}$ and Lorentz factor $\Gamma_{\rm f}$ encounters a slower counterpart characterized by $M_{\rm s}$ and $\Gamma_{\rm s}<\Gamma_{\rm f}$. The conversation of momentum gives the Lorentz factor of merged shell via the inelastic collision
\begin{equation}
    \Gamma_{\rm m}=\sqrt{\frac{M_{\rm f}\Gamma_{\rm f}+M_{\rm s}\Gamma_{\rm s}}{{M_{\rm f}/\Gamma_{\rm f}+M_{\rm s}/\Gamma_{\rm s}}}},
\end{equation}
for $\Gamma_{\rm f}>\Gamma_{\rm s}\gg1$, as derived in the Appendix \ref{app:collisions}. The energy dissipation can then be written as
\begin{equation}
    \mathcal E_{\rm is}=\left(M_{\rm f}\Gamma_{\rm f}+M_{\rm s}\Gamma_{\rm s}-M_{\rm m}\Gamma_{\rm m}\right)c^2,
\end{equation}
which implies the isotropic-equivalent energy density 
\begin{equation}
    u_{\rm in,iso}'=\frac{\mathcal E_{\rm is}}{\left({\theta_j^2}/{2}\right)\Gamma_{\rm m}V'_{\rm shell}},
\end{equation}
where $M_{\rm m}=M_{\rm f}+M_{\rm s}$ is the mass of the merged shell. Given the collision duration $\Delta T_{\rm coll}$ and the relative Lorentz factor $\Gamma_{\rm is,rel}=(\Gamma_{\rm f}/\Gamma_{\rm s}+\Gamma_{\rm s}/\Gamma_{\rm f})/2$ between the fast and slow shells, the comoving shell volume is $V'_{\rm shell}\sim4\pi R_{\rm coll}^2\Delta_{\rm shell}'$, where $R_{\rm coll}=\Gamma_{\rm m}^2c\Delta T_{\rm coll}/(1+z)$ denotes the collision radius and $\Delta_{\rm shell}'\sim\Gamma_{\rm is,rel}c\Delta T_{\rm coll}/(1+z)$ represents the comoving width of the merged shell. To fit the X-ray flare light curve and reduce the number of free parameters, while keeping the merged shell moving at the same velocity as the ejecta unshocked by the RS, we assume $M_{\rm f}=M_{\rm s}=M_{\rm shell}$ and fix $\Gamma_{\rm m}=\Gamma_0$, $\Delta T_{\rm coll}\sim\Delta T_{\rm X, flare}$, and $T_{\rm coll}\sim T_{\rm flare}\sim1.3~\rm d$ as the time at which the shell collision begins. Introducing $\epsilon_e^{\rm is}$, $\epsilon_B^{\rm is}$, and $f_e^{\rm is}$, one could parameterize the magnetic field strength and normalize the electron injection rate within the duration $t_{\rm is,inj}'=\Gamma_{\rm m}\Delta T_{\rm coll}/(1+z)$. 

Using the parameters listed in the lower part of Table~\ref{tab:params}, the green dash-dotted curve in the left panel of Figure~\ref{fig:lc_spec} shows the X-ray light curve produced by the internal shock scenario, which successfully reproduces the observed X-ray flare. Meanwhile, the synchrotron spectrum after absorption correction is consistent with the keV–MeV observations and is harder than the afterglow X-ray spectra, as illustrated in the right panel (solid green curve). Together, the light curve and spectral modeling support an internal shock origin for the X-ray flare, given that the isotropic-equivalent X-ray energy, $\mathcal E_{X,\rm iso}\sim \epsilon_e^{\rm is}\mathcal E_{\rm is}/(\theta_j^2/2)$ with $M_{\rm shell}=0.001M_\odot$, $\Gamma_{\rm f}\sim60$, and $\Gamma_{\rm s}=15$, is comparable to the observed flare energetics.

\begin{figure}
    \centering
    \includegraphics[width=1.0\linewidth]{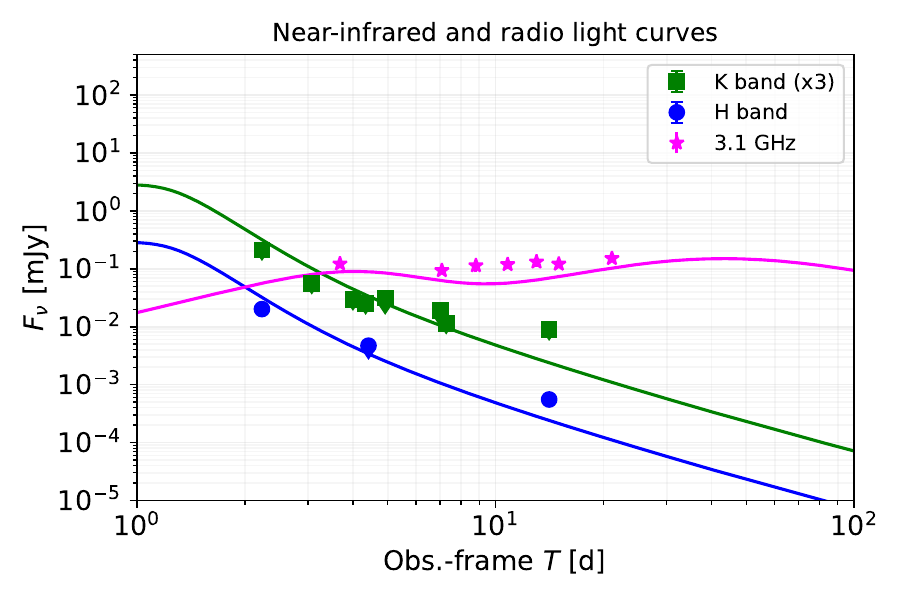}
    \caption{Near-infrared (H and K bands) and radio (3.1 GHz) light curves: external shock emission (solid curves) versus observations by Very Large Telescope (VLT) and MeerKAT \cite{Levan:2025mrt,Li:2025mae}.}
    \label{fig:multi_LCs}
\end{figure}

\section{Summary and Discussion}\label{sec:discussion}
Through modeling the time-dependent emission from a relativistic jet persistently powered by repeated partial disruptions of a WD by an IMBH, we demonstrate that the external shock regions, including both FS and RS, can account for the long-term X-ray afterglow of the ultra-long GRB 250702B over tens of days. Notably, the transition from a steep to a shallow decay at $T\sim10~\rm d$ can be naturally interpreted as the shift from RS to FS dominance. Moreover, we propose a phenomenological interpretation of the X-ray flare as arising from internal energy dissipation through collisions of relativistic shells, likely driven by the major peak of the accretion rate from the final complete disruption, as discussed quantitatively below. The jet scenario, incorporating both external and internal shocks, also successfully reproduces the keV–MeV spectra during the flaring and afterglow phases.

In addition to X-rays, the observed near-infrared (H and K bands) and radio (3.1 GHz) light curves after $T\sim2~\rm d$ can also be well described by the external shock model, as depicted in Figure \ref{fig:multi_LCs}. The contribution from internal shocks is not shown, as it fades very rapidly after the peak. The behavior of the near-infrared light curves is similar to that of the X-rays: the light-curve decay slopes become shallower after $T\sim10~\rm d$, whereas the two-hump feature of the radio light curve (magenta curve) originates from early-time RS and late-time FS radiation, which has also been observed in jetted TDE radio afterglows (e.g., Refs. \cite{Sato:2024zpq,Yuan:2024daj}).

The parameters used in this study are obtained through theoretical estimates and manual adjustments rather than detailed fitting via parameter scans. Nevertheless, these values are well aligned with theoretical expectations. For instance, the WD mass, the IMBH mass, and the initial value of $R_p$ are choose such that the final complete disruption time is temporally coincident with the time of the X-ray flare to be fitted in the left panel of Figure \ref{fig:lc_spec}. Meanwhile, the mass of the internal shells can be physically associated with the peak accretion rate $\dot M_{\rm BH,pk}$ at $t\sim3\times10^{4}~\rm s$, as we have
\begin{equation}
M_{\rm m}\sim\frac{\eta_j\dot M_{\rm BH,pk}\Delta T_{\rm X,flare}}{(1+z)}\left(\frac{1}{\Gamma_{\rm f}}+\frac{1}{\Gamma_{\rm s}}\right),
\end{equation}
which implies that {\bf the internal shell collisions are likely driven by fluctuations of the accretion rate, and the major peak of $\dot M_{\rm BH}$ leads to the major flare at $T\sim 1.3~\rm d$.} From this perspective, the early-time X-ray observations (e.g., $T\lesssim0.5~\rm d$) shown in the left panel of Figure \ref{fig:lc_spec}, which we do not attempt to reproduce in this work, could be attributed to prompt emissions from internal shell collisions driven by accretion peaks from partial disruptions, with appropriate corresponding $M_{\rm shell},~\Gamma_{\rm f}$, and $\Gamma_{\rm m}$. On the other hand, for external shocks, $n_{\rm ISM}$, $\theta_j$, $\eta_j\eta_{\rm acc}$, and $\Gamma_0$ are constrained by the energy normalization, as well as the slopes and the peak time of the X-ray light curve, since the jet deceleration is governed by $\int L_{j,\rm iso} dt \sim (4\pi/3)R_j^3\Gamma_{0}^2 n_{\rm ext} m_p c^2$ and the jet-break time is set by $\Gamma_{\rm sh}\theta_j=1$. In contrast, the microphysical parameters, such as $\epsilon_e$, $\epsilon_B$, and $f_e$, primarily determine the spectral shapes.

Above all, this study provides a framework for interpreting the temporal and spectral signatures of ultra-long GRBs from IMBH–WD TDEs, including long-lasting (days to weeks) X-ray, near-infrared, and radio afterglows, as well as late-time X-ray flares that reflect sustained central engine activity over hours to days. It also enables constraints on the tidal disruption and jet parameters.

\section*{Acknowledgments}
The work is partially funded by IISN project No. 4.4501.18, the National Natural Science Foundation of China (grant No. 12393812 and 12522303), and the Strategic Priority Research Program of the Chinese Academy of Sciences (grant NO. XDB0550300 and XDB0550200).  

\appendix
\section{Lorentz factor of the merged shell after relativistic shell collisions}\label{app:collisions}
We consider two spherical shells moving in the same direction and undergoing a perfectly inelastic collision. The mass, Lorentz factor, and velocity of the fast and slow shells are written as
\begin{equation}
\begin{split}
&\text{Fast shell:~} M_{\rm f},~\Gamma_{\rm f},~\beta_{\rm f}c;\\
&\text{Slow shell:~} M_{\rm s},~\Gamma_{\rm s},~\beta_{\rm s}c,\\
\end{split}
\end{equation}
with $\beta_{\rm f/s}=(1-\Gamma_{\rm f/s}^{-2})^{1/2}$. In the IMBH-rest frame, the total energy and momentum before the collision are $E_{\rm tot}=(\Gamma_{\rm f}M_{\rm f}+\Gamma_{\rm s}M_{\rm s})c^2$ and $P_{\rm tot}=(\Gamma_{\rm f}M_{\rm f}\beta_{\rm f}+\Gamma_{\rm s}M_{\rm s}\beta_{\rm s})c$. The effective mass of the merged shell is determined by the invariant $M_{\rm eff}^2c^4=E_{\rm tot}^2-P_{\rm tot}^2c^2$, which yields the bulk Lorentz factor
\begin{equation}\begin{split}
\Gamma_{\rm m}^2&=\frac{E_{\rm tot}^2}{M_{\rm eff}^2c^4}\\
&=\frac{(\Gamma_{\rm f}M_{\rm f}+\Gamma_{\rm s}M_{\rm s})^2}{(\Gamma_{\rm f}M_{\rm f}+\Gamma_{\rm s}M_{\rm s})\left(\frac{M_{\rm f}}{\Gamma_{\rm f}}+\frac{M_{\rm s}}{\Gamma_{\rm s}}\right)+\Lambda},
\end{split}
\label{eq:gamma_m2}
\end{equation}
where
\begin{equation}\begin{split}
\Lambda &= M_{\rm f}M_{\rm s}\Gamma_{\rm f}\Gamma_{\rm s}(\beta_{\rm f}-\beta_{\rm s})^2\\
&= M_{\rm f}M_{\rm s}\Gamma_{\rm f}\Gamma_{\rm s}\left(\frac{1}{4\Gamma_{\rm f}^4}+\frac{1}{4\Gamma_{\rm s}^4}+\mathcal O(\Gamma_{\rm f}^{-6}+\Gamma_{\rm s}^{-6})\right),
\end{split}
\end{equation}
is an infinitesimal quantity compared to the first term in the denominator of Eq. \ref{eq:gamma_m2} for $\Gamma_{\rm f}>\Gamma_{s}\gg1$. We then obtain a simple expression for $\Gamma_{\rm m}$,
\begin{equation}
\Gamma_{\rm m}\approx\sqrt{\frac{M_{\rm f}\Gamma_{\rm f}+M_{\rm s}\Gamma_{\rm s}}{{M_{\rm f}/\Gamma_{\rm f}+M_{\rm s}/\Gamma_{\rm s}}}}.
\end{equation}




\bibliographystyle{apsrev4-2}
\bibliography{ref}

\end{document}